# A Hybrid Density Functional Study of Armchair Si and Ge Nanotubes

Prachi Pradhan and Asok K. Ray*
Department of Physics
The University of Texas at Arlington
P.O Box 19059
Arlington, Texas 76019

*akr@uta.edu

# ABSTRACT

First principles calculations based on hybrid density functional theory have been used to study the electronic and geometric properties of armchair silicon and germanium nanotubes ranging from A (3, 3) through A (9, 9). The approach used is the finite cluster approach with hydrogen termination to simulate the effects of longer tubes. A detailed comparison of the structures and stabilities of Si and Ge nanotubes has been performed and the dependence of the HOMO- LUMO or "band" gaps on the tube diameters has been investigated. Silicon nanotubes appear to be "less-puckered" and more stable compared to germanium nanotubes. The largest silicon nanotube studied has a cohesive energy of 3.138eV/atom to be compared with the cohesive energy of 2.770eV/atom for the corresponding germanium nanotube. Contrary to some published results in the literature, silicon nanotubes do *not* appear to be metallic for the cases studied in the armchair configuration.

## 1. Introduction

Discovery and synthesis of the "buckyball" $C_{60}$ in the early 1990's led scientists towards a new miracle material called *carbon nanotubes* (CNTs) which are thin tubes of carbon atoms.[1,2] With the tremendous success of CNTs, over the years quasi-one-dimensional nanostructures such as nanotubes and nanowires have stirred extensive interests in condensed matter physics and in fact, the entire research world, partly because of their fascinating electronic and mechanical properties and partly because of novel technological applications.[3-11] Single walled nanotubes particularly have been studied more extensively both experimentally and theoretically. One interesting fact observed both experimentally and theoretically in the case of carbon nanotubes is that single walled carbon nanotubes are believed to exhibit metallic or semiconducting behavior depending on the tube diameter and chirality. Length and curvature also are found to influence the structure and energetics of a nanotube.[12-16]

Continuing the extensive studies and applications of CNTs, synthesis of several other nanotubes with different materials have been reported, for example, $NiCl$, $NiCl_2$, $H_2Ti_3O_3$, and $TiO_2$.[17-19] Theoretical studies on GaN, GaSe, P, and Cu nanotubes have also been reported.[20-23] In this area, silicon, the workhorse of semiconductor industry, occupies a central place. Research in recent years have underscored the importance and development of silicon at nanoscale. Nano-silicon is believed to be a potential candidate for diverse applications such as creating better disease detectors and biochemical sensors, as well as tiny electronics such as ultra-high density memory chips for ultra fast computing. Silicon nanotubes have received much attention, followed by two successful attempts to synthesize them. Sha *et al.*[24] prepared Si nanotubes by chemical vapor deposition (CVD) process using nanochannel $Al_2O_3$ (NCA) substrate. Recently, Jeong *et al.*[25] reported synthesis of SiNTs on porous alumina using molecular beam epitaxy (MBE).

Following this line of advances in the periodic table, we note that germanium is an element in the same column IV of the periodic table just below silicon. As known already, there is a stark similarity between bulk silicon and germanium regarding their structural and electronic properties. They both usually form diamond-like 3D structures with tetrahedrally coordinated $sp^3$ hybridized atoms. Studies also have shown similarities between germanium and silicon clusters supported by experimental results.[26,27] In principle, since carbon nanotubes exist, there is no academic reason to doubt the existence of silicon and germanium nanotubes. We do hasten to point out that *no* experimental detection of germanium nanotubes has been reported *yet*. Part of the reason for CNTs is attributed to $sp^2$ bonding preferred by carbon. But, if graphene-like sheets of silicon and germanium can be formed, rolled nanotubes are possible. Contrary to the popular belief that nanotubes of elemental silicon or germanium are difficult to stabilize owing to the preference for $sp^3$ hybridization these tubes can, in principle, be stabilized by proper termination of dangling bonds on the open ends of the tubes.

There are, in general, two approaches for constructions of nanotubes. One approach is called the "cluster-stacking approach" in which member rings of Si or Ge are stacked on top of each other to form a tubular structure and another approach is "cluster-CNT based", which comprises of rolling a graphene-like sheet of Si or Ge to form a tube. Seifert *et al.*[28] proposed silicon based tubular nanostructures by rolling the (Si-H) silicide sheets to form Si-H nanotubes. They concluded that, silicide and SiH nanotubes have semi-conducting gaps, independent of chirality, which converged rapidly with increasing diameter to that of the 2D layer. Based on stability considerations compared to CNT's, they argued that the synthesis of silicide as well as SiHNTs could be achieved. Fagan *et al.*[29] did a comparative LDA-DFT (local density approximation to density functional theory) study of the electronic, the structural, and thermal properties of three infinite silicon nanotubes (SiNTs), namely armchair (6, 6), zigzag (6, 0), and mixed chiral (8, 2) structures. They found that, similar to carbon nanotubes, silicon nanotubes may present metallic (armchair) or semiconductor (zigzag and mixed) behaviors, depending on their chiralities. The gap was found to decrease in inverse proportion to the diameter. They also performed a Monte-Carlo simulation study using the Tersoff''s potential and found the existence of relevant discrepancies regarding the thermal stabilities and energy differences between cohesive energies per atom for C and Si tubes compared with their corresponding bulks. Zhang *et al.*[30] used the semi-empirical molecular orbital theory PM3 and Hartree-Fock (HF) theory with two different basis sets, namely 3-21G and 3-21G(d), to study four models, a silicon nanowire (SiNW), a silicon nanotube (SiNT), a carbon nanowire (CNW) and a carbon nanotube (CNT). These studies showed that at the HF level, results are very sensitive to the basis set used and silicon tubular structures are less stable compared to their carbon counterparts. However under appropriate conditions silicon nanotubes with puckered surface structures are possible. Zhang *et al.*[31] studied three structures of finite SiNT's at B3LYP/6-31G (d) level. Their study predicts that the armchair silicon structures are more stable compared to the zigzag structures. Barnard and Russo[32] have investigated the energetics and structural properties of infinite armchair and zigzag silicon nanotubes (SiNT) as a function of tube diameter. They studied A(3,3) through A(9,9) and Z(3,0) through Z(9,0) nanotubes. All calculations were performed with the Vienna *ab initio* simulation package (VASP) using the generalized gradient approximation to density functional theory (GGA-DFT) with the Perdew and Wang functional. Ultra-soft, gradient-corrected-Vanderbilt-type pseudopotentials were used and the studies showed that the atomic heat formation of a silicon nanotube is dependent on the nanotube diameter, but independent of the chiral structure of the tube. Also, it was shown that the individual cohesive energies and strain energies are dependent on both diameter and chirality. The response of hypothetical silicon nanotubes under axial compression has been investigated by Kang *et al*[33]. using atomistic simulations based on the Tersoff potential. Bai *et al.*[34] have used molecular dynamics and cluster-stacking approach to predict the possible existence of one-dimensional silicon nanostructures: the square, pentagonal, and hexagonal single-walled silicon nanotubes. They concluded that the stacked short and thin nanotubes were locally stable in vacuum and have zero band gap, suggesting that the SWSiNTs are possibly metals rather than wide-gap semiconductors. Using *ab initio* total energy calculations, Singh *et al.* have studied metal encapsulated nanotubes of silicon.[35] They found the finite nanotubes to have varying HOMO-LUMO gaps depending on the length and amount of doping but infinite nanotubes were metallic, symmetric, and stable. There has been only one study on Ge-H nanotubes, a density-functional tight binding study by Seifert *et al.*[36] where the tubes were constructed by rolling sheets of polygermyne. These results indicate a possibility of germanium nanotubes with interesting photoluminescence properties and the GeH nanotubes were found to be semi-conducting, with the gap size growing from about 1.1eV from the smallest nanotube towards the value of 1.35eV of germyne sheets. All these studies indicate that significant controversies exist about silicon nanotubes. *Also, as far as we know, there has been no study up till now using the cluster-CNT based approach to study bare germanium nanotubes. Ours is the first attempt to study both silicon and germanium nanotubes on an equal basis from this perspective.* We examine below silicon and germanium nanotubes with varying sizes and present a comparative electronic structure analysis between silicon and germanium nanotubes.

## 2. Results and discussions

Our approach for construction of the nanotubes is based on single-walled carbon nanotubes and the finite cluster approach. The easiest way to visualize how nanotubes are built is to start with graphite-like sheets of silicon and germanium. Then a single–walled nanotube is constructed by wrapping one single layer of the graphite‐like sheet to form a cylindrical shape. Starting with only one layer of 2-dimensional graphite-like sheet we end up with a cylinder with only one wall, namely a single wall nanotube (SWNT). If more layers are taken into account, cylinders with multiple walls may result, a multi wall nanotube (MWNT). Our interest here lies only in the single-walled nanotube. The structure of such nanotubes may be described in terms of chirality and length. Chirality and diameter are specified in terms of magnitude of the components of chiral vector. The vector $C_h$ which maps an atom from the left hand border onto an atom on the right border line is an integer multiple of the two basis vectors $a_1$ and $a_2$, i.e., $C_h = n\ a_1 + m\ a_2$ with integers n and m. Depending upon how the sheet is rolled we have three types of tubes. For armchair m = n, for zigzag m = 0 and for chiral nanotubes m ? n. In this work, we have concentrated on armchair SWNTs.

As far as computations are concerned, we note that given the large sizes of the systems under consideration, an accurate method with low computational overhead is both necessary and desirable. The two standard methods in computational condensed matter physics, one based on Hartree-Fock (HF) theory and the other on density functional theory (DFT) in the local density approximation (LDA) or in the generalized gradient approximation (GGA), both have their advantages and disadvantages.[37, 38] On the other hand, hybrid density functional theory incorporating Hartree-Fock exchange with DFT exchange-correlation have shown to be a promising method. Therefore, we have used B3LYP[39], a hybrid functional, and the Los Alamos pseudopotential LANL2DZ[40] with the associated basis set (for Si atom the electrons 1s 2s 2p orbitals and for Ge the electrons in 1s 2s 2p 3s 3p 3d orbitals are replaced by core potentials) to perform atomistic simulations of silicon and germanium nanotubes. The quality of B3LYP/LANL2DZ scheme for the description of Si and Ge nanotubes was tested by calculations on Si and Ge atoms and the dimers. For Si atom, the ionization potential and electron affinity are 8.462eV and 0.896eV to be compared with the experimental values of 8.151eV and 1.385eV, respectively. For Ge atom, our values are 8.008eV and 0.841eV, and the experimental values are 7.9eV and 1.233eV, respectively. A very large basis set is necessary for the theoretical electron affinity to approach the value of the experimental electron affinity and this was not considered to be necessary and computationally feasible for the large clusters representing the nanotubes studied in this work. For Si and Ge dimers, our bond lengths are 2.352 Å and 2.527 Å, to be compared with the experimental values of 2.32 Å and 2.44 Å, respectively. For SiH and GeH, our values are 1.547 Å and 1.630 Å, respectively and the experimental values are 1.52 Å and 1.59 Å.[41] Thus, overall, our choices of the exchange-correlation potential and the pseudopotential with the associated basis set can be considered to be quite satisfactory. The cohesive energy or the binding energy per atom for each system was computed as per the following formula:

$$E_b = \{[mE(X) + nE(H)] - [E(X_m H_n)]\} / (m+n) \qquad (1)$$

where m is the number of silicon or germanium atoms and n is the number of hydrogen atoms in the nanotube, and E(X) where (X=Si or Ge) and E(H) are the ground state total energies of the silicon or germanium and hydrogen atoms respectively. $E(X_m H_n)$ is the total energy of the optimized clusters representing the nanotubes.

In figures 1-2, we present the structures of silicon and germanium nanotubes. Figure 3 gives a detailed description of the optimized bond lengths (Angstroms), bond angles and co-ordination between the atoms for a sample Si (4, 4) and Ge (4, 4) nanotube. All structures are Berny geometry and spin-optimized.[42] In tables 1-2 we present results for the comparative study of the electronic and geometric structures of Si and Ge nanotubes in armchair A (3, 3) through A (9, 9) structures. The HOMO-LUMO gap is computed as the energy difference between the highest occupied molecular orbital and the lowest unoccupied molecular orbital. All computations have been performed using the *Gaussian '03* suite of programs[43] at the Supercomputing Center of the University of Texas at Arlington.

As mentioned before, all seven systems for Si or Ge have been modeled using the finite-cluster-CNT based approach. The smallest armchair silicon nanotube Si (3, 3) SiNT is modeled as $Si_{60}H_{12}$ cluster. The largest system studied for silicon nanotubes SiNT is Si (9, 9) nanotube modeled as $Si_{180}H_{36}$. For germanium nanotube, the corresponding clusters are $Ge_{60}H_{12}$ and $Ge_{180}H_{36}$. Hydrogen termination is used for the nanotubes to simulate the effect of longer tubes and saturate the dangling bonds. The initial configurations of the nanotubes were based on the assumption that the nanotubes can be constructed by folding a 2D graphene like sheets of Si/Ge with a bond distance of 2.25 ? for Si and 2.45 ? for Ge.

As shown in figures 1 and 2 the optimized structures of the silicon and germanium nanotubes show a puckered or corrugated appearance. The smallest tubes armchair Si (3, 3) and Ge (3, 3) show a slightly more non-uniform diameter than other configurations. With the increase in tube diameters from A (3, 3) to A (9, 9) the structures tend to be more uniform. As mentioned before, figure 3 shows the coordination of Si and Ge atoms in the armchair Si (4,4) and Ge(4,4) nanotubes in the middle part of the tube. For SiNT (4, 4) (Figure 3a) the silicon atoms are not in same plane with Si-Si-Si angles of 117º and 119º, respectively. The Si-Si bond lengths alternate between 2.237 ? to 2.281 ?, close to the single bond values. However, on the periphery of the tube the hexagonal rings appear to show more alternation in bond lengths, between 2.188 ? (Si = Si) to 2.468 ? (Si – Si). This bond length alternation of 0.28 ? is more pronounced than that in CNTs and exhibits a strong tendency for bond localization. This trend continues for the other tubes too.

In the case of GeNT (4, 4) (Figure 3b) the Ge-Ge bond lengths vary from 2.383 to 2.451 ?, throughout the tube length, showing a bond alternation of 0.068 ?. This is larger than the alternation observed the middle part of the SiNT (4, 4) (0.044 ? ). Poor p-p overlaps, hence weak p bonding, suggests more bond alternation, i.e. less electron delocalization. In case of GeNTs this is more pronounced than SiNTs. Hence, we see a more corrugated tubular shape for GeNTs than that for SiNTs.

Table 2 and figure 4 show the variations of the cohesive energies per atom in eV versus the number of atoms in the Si/Ge nanotubes. As the number of atoms increase the cohesive energy of silicon nanotubes increases. The largest SiNT studied Si (9, 9) has a cohesive energy of 3.138eV/atom, about 68% of the bulk cohesive energy of 4.63 eV/atom. The cohesive energies for germanium are lower than that of silicon as expected and show an increasing trend although not a smooth one as SiNT. The largest GeNT studied Ge (9, 9) has a cohesive energy of 2.770 eV/ atom, about 72% of the bulk cohesive energy of 3.85eV/atom. As a comparison, the cohesive energy of carbon nanotube can be as high as 99% of the bulk. This does *not* however necessarily rule out the existence of SiNT and GeNTs.

One of the central questions in the theory and applications of nanotubes is the possible metallic or semi-conducting properties of these tubes. To examine this we calculated the highest-occupied-molecular-orbital to lowest-unoccupied-molecular-orbital (HOMO-LUMO) gap. These can provide a measure of the band gap for the infinite solid as the number of atoms in the cluster increases and also to analyze the conductivity of the nanotube. Table 3 and figures 5a and 5b show the variation of the HOMO-LUMO gaps with the tube diameter for silicon and germanium nanotubes respectively. The gaps for the silicon nanotubes are in the range of 0.885eV to 1.023eV. These gaps are smaller than the bulk silicon gap of 1.1eV but still do *not* indicate any metallic behavior of silicon nanotubes even for the largest cluster studied. We do note that as we go beyond the Si (6, 6) nanotube with the tube diameter increasing, the gap decreases. This feature is different from that observed in case of carbon nanotubes, which were found to be metallic in armchair configuration. In case of germanium there is no definite pattern with respect to the increase in tube diameter and the gaps show an oscillatory pattern. The range for the gaps is between 0.274ev -0.865eV. For most of the Ge nanotubes, the gaps are significantly lower that the bulk value of 0.7eV. In fact, A (5, 5) and A (6, 6) Ge nanotubes have gaps of only 0.274 and 0.252eV and it is *possible* that Ge nanotubes can exhibit metallic characteristics depending on the tube diameter. For both materials in armchair configuration, the gaps do appear to be decreasing indicating a *possible* metallic behavior in the infinite limit.  The high ratio of $sp^3$ to $sp^2$ hybridization and the extent of overlap of the p and s bonding clearly contribute to a different behavior for Si and Ge nanotubes as compared C nanotubes. In any case, our results indicate that germanium nanotubes have more of a metallic character than that of silicon nanotubes.

In conclusion, we have examined silicon and germanium nanotubes in armchair configurations. Our results show that the germanium nanotubes are possible, though from the point of view of stability, silicon nanotubes appear to be more stable compared to germanium nanotubes. Also the germanium nanotubes due to the higher $sp^3$ character appear to be more puckered in appearance than their carbon and silicon counterparts. In our previous work on mixed silicon carbon clusters, we found that with proper stoichiometery of silicon and carbon atoms in a cluster, one can achieve highly stable clusters [44]. One way to improve the $sp^2$ character of Si and Ge tubes and thus possibly increase stability would be to add dopant atoms like carbon.  Research is underway to study various types of doping like substitutional or interstitial doping and study the characteristics and stabilities of these systems.

Finally, the authors gratefully acknowledge partial support from the Welch Foundation, Houston, Texas (Grant No. Y-1525).

**References**

[1] S. Iijima, Nature **354**, 56 (1991); S. Iijima and T. Ichihashi, Nature 363, 603 (1993).

[2] J. W. Mintmire, B. I. Dunlap, and C. T. White, Phys. Rev. Lett. 68**,** 631 (1992).

[3] W. Liang, M. Bockrath, D. Bozovic, J. H. Hafner, M. Tinkham, and H. Park, Nature 411, 665 (2001).

[4] S. P. Franh, P. Poncharal, Z. L. Wang, and W. A. de Heer, Science 280**,** 1744 (1998).

[5] M. J. Biercuk, M. C. Llaguno, M. Radosavljevic, J. K. Hyun, A. T. Johnson, and J. E. Fischer, Appl. Phys. Lett. 80, 2767 (2002).

[6] Y. Chen, D. T. Shaw, X. D. Bai, E. G. Wang, C. Lund, W. M. Lu, and D. D. L. Chung, Appl. Phys. Lett. 78, 2128 (2001).

[7] A. C. Dillon and M. J. Heben, Appl. Phys. A. 72, 133 (2001).

[8] N. S. Lee, D. S. Chung, I. T. Han, J. H. Kang, Y. S. Choi, H. Y. Kim, S. H. Park, Y. W. Jin, W. K. Yi, M. J. Yun, J. E. Jung, C. J. Lee, J. H. You, S. H. Jo, C. G. Lee, and J. M. Kim, Diamond Relat. Mater. 10, 265 (2001).

[9] H. Sugie, M. Tanemura, V. Filip, K. Iwata, K. Takahashi, and F. Okuyama, Appl. Phys. Lett. 78, 2578 (2001).

[10] C. L. Cheung, J. H. Hafner, T. W. Odom, K. Kim, and C. M. Lieber, Appl. Phys. Lett. 76, 3136 (2000).

[11] P. G. Collins, K. Bradley, M. Ishigami, and A. Zett, Science 287, 1801 (2000).

[12] N. Hamada, S. I. Sawada, and A. Oshiyama, Phys. Rev. Lett. 68, 1579 (1992).

[13] R. Saito, M. Fujiata, G. Dresselhaus, and M. S. Dresselhaus, App. Phys. Lett. 60, 2204 (1992).

[14] T. W. Odom, J. L. Huang, P. Kim, C. M. Lieber, Nature 391**,** 62 (1998).

[15] R. A. Jishi, J. Bragin, and L. Lou, Phys. Rev. B 59, 9852 (1999).

[16] O. Gulseren, T. Yildirim, S. Ciraci, Phys. Rev. Lett. 65, 3405 (2002).

[17] Y. R. Hacohen, E. Grunbaum, R. Tenne , J. Sloand, and J. L. Hutchinson, Nature 395, 336 (1998); Y. R. Hacohen, R. Popovitz-Biro, E. Grunbaum, Y. Prior, and R. Tenne, Adv. Mater. 14, 1075 (2002).

[18] Q. Chen , W. Zhou , G. Du, and L. M. Peng , Adv. Mater. 14, 1208 (2002).

[19] G. R. Patzke, F. Krumeich, and R. Nesper, Angew. Chem. Int. Ed. 41, 2446 (2002).

[20] S. M. Lee, Y. H. Lee, Y. G. Hwang, J. Elsner, D. Porezag, and Th. Frauenheim, Phys. Rev B 60, 7788 (1999).

[21] M. Cote, M. L. Cohen, and D. J. Chadi, Phys. Rev. B. 58, 4277 (1998).

[22] G. Seifert and E. Hernandez, Chem. Phys. Lett. 318, 355 (2000).

[23] J. W. Kang, H. J. Hwang, and J. J. Seo, J. Phys. Cond. Matt. 14, 8997 (2002).

[24] J. Sha, J. Niu, X. Ma, J. Xu, X. Zhang, Q. Yang, and D. Yang Adv. Mater. 14, 1219 (2002).

[25] S. Jeong, J. Kim, H. Yang, B. Yoon, S. Choi, H. Kang, C. Yang, and Y. Lee, Adv. Mat. 15 , 1172 (2004).

[26] B. X. Li, M. Jiang, and P. L. Cao, J. Phys. Cond. Matt. 11, 8517(1999); D. A. Dixon, and J. L. Gole, Chem. Phys. Lett. 188, 560 (1992).

[27] J. M. Alford, R. T. Laaksonen, and R. E. Smalley, J. Chem. Phys. 94, 2618 (1991).

[28] G. Seifert, Th. Kohler, H. M. Urbassek, E. Hernandez, and Th. Frauenheim, Phys. Rev. B 63, 193409 (2001).

[29] S. B. Fagan, R. J. Baierle, R. Mota, A. J. R. da Silva, and A. Fazzio, Phys. Rev. B 61, 9994 (2000); S. B. Fagan, R. Mota, R. J. Baierle, G. Paiva, A. J. R. da Silva, and A. Fazzio, J. Mol. Struct. 539, 101 (2001).

[30] R. Q. Zhang, S. T. Lee, C.- K. Law, W.-K. Li, and B. K. Teo, Chem. Phys. Lett. 364, 251 (2002); Chem. Phys. Lett. 368, 509 (2003).

[31] M. Zhang, Y. H. Kan, Q. J. Zang, Z. M. Su, and R. S. Wang, Chem. Phys. Lett. 379, 81 (2003).

[32] A. S. Barnard and S. P. Russo, J. Phys. Chem. 107, 7577 (2003).

[33] J. W. Kang and H. J. Hwang, Nanotechnology 14, 402 (2003); J. W. Kang , K. R. Byun, and H. J. Hwang, Modelling Simul. Mater. Sci. Eng. 12, 1 (2004).

[34] J. Bai, X. C. Zeng, H. Tanaka, and J. Y. Zeng, Proc. Nat. Acad. Sci. 101, 2664 (2004).

[35] A. K. Singh, V. Kumar, T. M. Briere, and Y. Kawazoe, Nano Lett. 2, 1243 (2002); A. K. Singh, T. M. Briere, V. Kumar, and Y. Kawazoe, Phys. Rev. Lett. 91, 146802 (2003); A. K. Singh, V. Kumar, and Y. Kawazoe, J. Mater. Chem. 14, 555 (2004).

[36] G. Seifert, Th. Kohler, Z. Hajnal, Th. Frauenheim, Sol. State Comm. 119, 653 (2001).

[37] W. J. Hehre, L. Radom, P. v. R. Schleyer, and J. A. Pople, *Ab Initio Molecular Orbital Theory* (Wiley, New York, 1986); D. C. Young, *Computational Chemistry*, (Wiley, New York, 2001).

[38] P. Hohenberg and W. Kohn, Phys. Rev. 136, B864 (1964); W. Kohn and L. J. Sham, Phys Rev. 140, A1133 (1965); D. M. Ceperley and B. J. Adler, Phys. Rev. Lett. 45, 566 (1980); J. C. Slater, *Quantum Theory of Molecules and Solids, Vol. 4: The Self-*

*Consistent-Field for Molecules and Solids* (McGraw-Hill, New York, 1974); S. H. Vosko, L. Wilk, and M. Nusair, Can. J. Phys. 58, 1200 (1980); R. G. Parr and W. Yang, *Density Functional Theory of Atoms and Molecules* (Oxford University Press, New York, 1989); R. O. Jones and O. Gunnarsson, Rev. Mod. Phys. 61, 689 (1989); J. P. Perdew, in *Electronic Structure of Solids*, Ed P. Ziesche and H. Eschrig (Akademie Verlag, Berlin, 1991); J. P. Perdew, S. Kurth, A. Zupan and P. Blaha, Phys. Rev. Lett. 82, 2544 (1999); J. F. Dobson and J. Wang, Phys. Rev. B 62, 10038 (2000).

[39] A. D. Becke, J. Chem. Phys. 98, 5648 (1993); J. Chem. Phys. 109, 2092 (1998); C. Lee, W. Yang, and R. G. Parr, Phys. Rev. B 37, 785 (1988).

[40] P. J. Hay and W. R. Wadt, J. Chem. Phys. 82, 270 (1995).

[41] *CRC Handbook of Chemistry and Physics,* 76[th] Edition (CRC Press, 1995).

[42] J. B. Foresman and AE. Frisch, *Exploring Chemistry with Electronic Structure Methods,* 2[nd] ed. (Gaussian, Inc. Pittsburgh, PA, 1996).

[43] *Gaussian 03*, Revision A.1, M. J. Frisch *et al.*, Gaussian Inc. Pittsburgh, Pa, 2003.

[44] M. N. Huda and A. K. Ray, Phys. Rev. A (Rapid Comm.) 69, 011201 (2004); Eur. Phys. J. D 31, 63 (2004); P. Pradhan and A. K. Ray, J. Mol. Struc. (Theochem) 716, 109 (2004); Bull. Am. Phys. Soc. 50, 1169 (2005); A. Srinivasan, M. N. Huda, and A. K. Ray, Bull. Am. Phys. Soc. 50, 302 (2005).

**Table 1.** $E_c$: Cohesive energies / atom (in eV) for the armchair nanotubes of silicon and germanium.

| Nanotube | Model | $E_{c-SiNT}$ | $E_{c-GeNT}$ |
|----------|-------|--------------|--------------|
| A (3,3) | $X_{60}H_{12}$ | 2.969 | 2.412 |
| A (4,4) | $X_{80}H_{16}$ | 3.056 | 2.456 |
| A (5,5) | $X_{100}H_{20}$ | 3.09 | 2.477 |
| A (6,6) | $X_{120}H_{24}$ | 3.113 | 2.490 |
| A (7,7) | $X_{140}H_{28}$ | 3.125 | 2.571 |
| A (8,8) | $X_{160}H_{32}$ | 3.133 | 2.512 |
| A (9,9) | $X_{180}H_{36}$ | 3.138 | 2.770 |

**Table 2.** Tube diameters $d_o$ in Å and HOMO-LUMO gaps in eV for the armchair nanotubes of silicon and germanium.

| Nanotube | $d_{o\ SiNT}$ | Gap $_{SiNT}$ | $d_{o\ GeNT}$ | Gap $_{GeNT}$ |
|----------|---------------|---------------|---------------|---------------|
| A (3,3) | 6.26 | 0.885 | 6.485 | 0.865 |
| A (4,4) | 8.389 | 0.944 | 8.876 | 0.855 |
| A (5,5) | 10.58 | 1.023 | 11.161 | 0.274 |
| A (6,6) | 12.78 | 1.067 | 13.601 | 0.252 |
| A (7,7) | 14.95 | 1.054 | 15.44 | 0.766 |
| A (8,8) | 17.041 | 1.019 | 18.402 | 0.633 |
| A (9,9) | 19.22 | 0.993 | 20.674 | 0.628 |

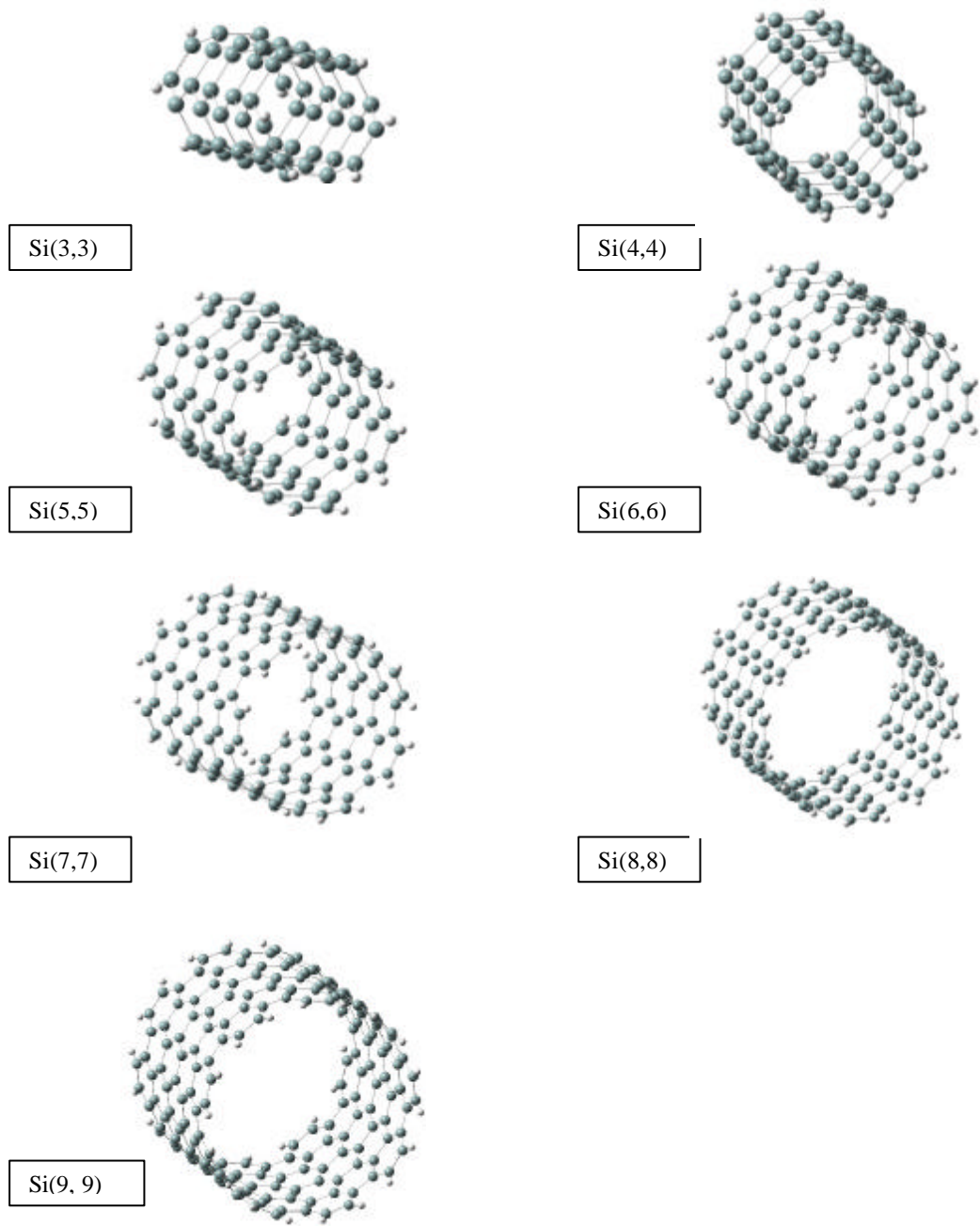

Figure 1. Structures of armchair silicon nanotubes from Si(3, 3) to Si (9, 9)

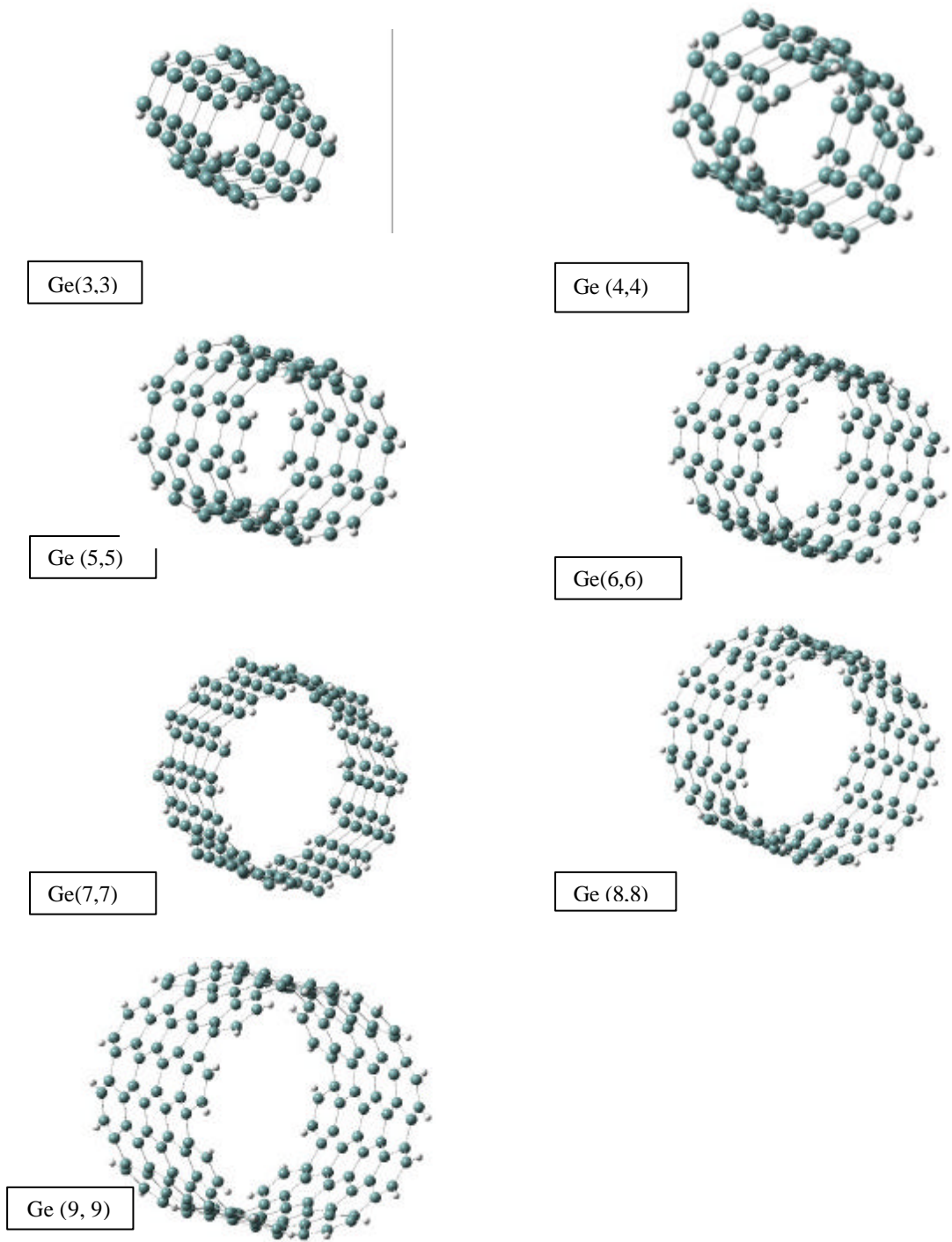

Figure 2. Structures of armchair germanium nanotubes from Ge (3, 3) to Ge (9,9)

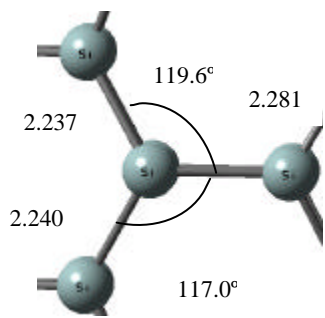

3(a)

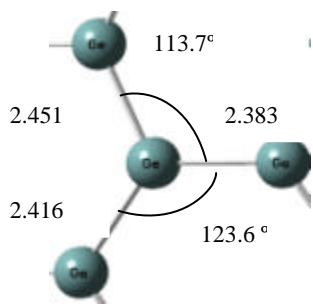

3(b)

Figure 3. The optimized bond lengths (? ) and angles (º) for 3(a) Si(4,4) silicon and 3(b) Ge(4,4) germanium nanotubes.

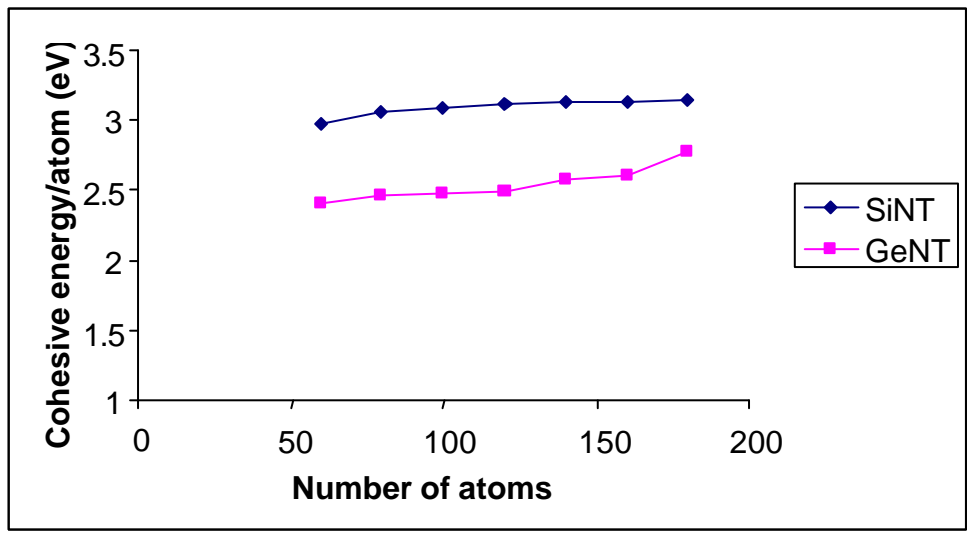

Figure 4. Cohesive energy/atom (eV) vs. number of silicon and germanium atoms in armchair nanotubes.

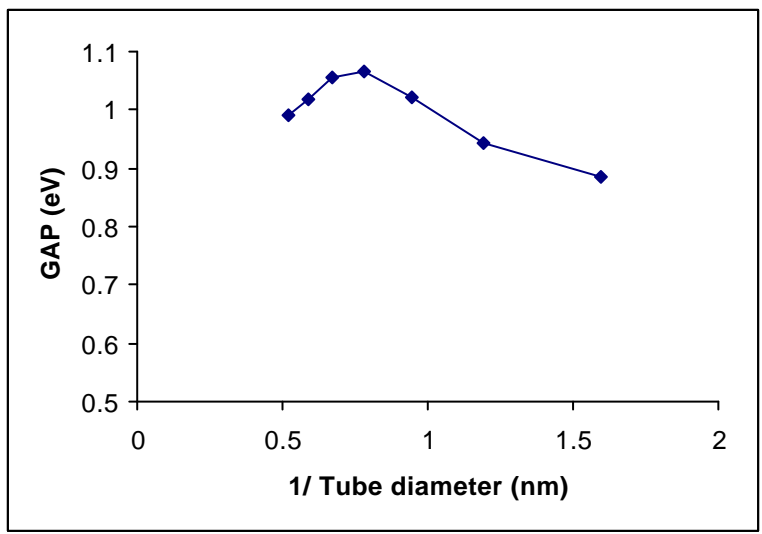

(5a)

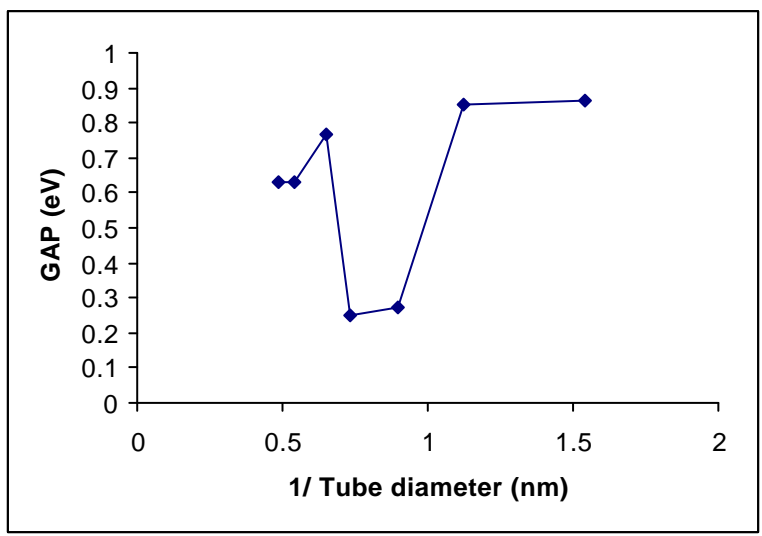

(5b)

Figure 5.  HOMO-LUMO gap in eV vs. 1/tube diameter in nm$^{-1}$ for (a) silicon armchair nanotubes and (b) germanium armchair nanotubes.